\documentclass[aps,notitlepage,a4paper,superscriptaddress,12pt]{extarticle}

\usepackage{multicol}
\usepackage[utf8]{inputenc}
\usepackage[dvips]{color}
\usepackage[toc]{appendix}
\usepackage[hang,flushmargin]{footmisc} 
\usepackage{titlesec,titletoc,ntheorem-hyper,authblk,abstract,latexsym,microtype,booktabs,amsmath,hyperref,cleveref,fancyhdr,wrapfig,array,amsfonts, amssymb,geometry,appendix,cite, secdot, tensor}
\usepackage[dvipsnames]{xcolor}

\newcommand{\be}{\begin{equation}}
\newcommand{\ee}{\end{equation}}
\newcommand{\bea}{\begin{eqnarray}}
\newcommand{\eea}{\end{eqnarray}}
\newcommand\blfootnote[1]{%
  \begingroup
  \renewcommand\thefootnote{}\footnote{#1}%
  \addtocounter{footnote}{-1}%
  \endgroup
}

\usepackage{graphicx,lipsum, cuted}
\usepackage{caption}
\usepackage{subfigure}
\usepackage{amsmath}
\usepackage{amsfonts}
\usepackage{amssymb}
\usepackage{stackengine}
\usepackage{mathtools}
\numberwithin{equation}{section}
\usepackage{titlesec}

\numberwithin{subcase}{case}
\usepackage{framed}
\allowdisplaybreaks
\usepackage[nottoc]{tocbibind}
\usepackage[free-standing-units]{siunitx}
\usepackage{helvet}
\usepackage{url}
\usepackage{titletoc}
\usepackage{blindtext}
\usepackage[gen]{eurosym}

\title{Analogue Hawking radiation as a tunneling in a two-level $\mathcal{PT}$-symmetric system}
\author{Bijan Bagchi$^{1,\dagger,*}$, Rahul Ghosh$^{2,\dagger}$, and Sauvik Sen$^{3,\dagger}$}
 \affil{Department of Physics, Shiv Nadar Institution of Eminence,\\ Gautam Buddha Nagar, Uttar Pradesh 203207, India}
\date{\today}

\begin{document}
\maketitle

\begin{abstract}
   In the light of a general scenario of a two-level non-Hermitian $\mathcal{PT}$-symmetric Hamiltonian we apply the tetrad-based method to analyze the possibility of analogue Hawking radiation. It is done by making use of the conventional null-geodesic approach wherein the associated Hawking radiation is described as a quantum tunneling process across a classically forbidden barrier which the event horizon imposes. An interesting aspect of our result is that our estimate for the tunneling probability is independent of the non-Hermitian parameter that defines the guiding Hamiltonian.
    \end{abstract}

\blfootnote{e-mail : 1. bbagchi123@gmail.com, 2. rg928@snu.edu.in, 3. sauviksen.physics@gmail.com}
\blfootnote{$*$ : corresponding author}
\blfootnote{$\dagger$ : These authors contributed equally to this work.}

{Keywords: Hawking radiation, $\mathcal{PT}$-symmetry, tunneling probability, exceptional points}\\
\maketitle

\section{Introduction}

The physics of black holes has continually aroused interest after Bekenstien-Hawking's pioneering works in the 1970s in trying to interpret them as thermodynamical objects which release radiation outside their event horizon \cite{bek1, bek2, hawk1, hawk2} (see, for a review of literature, \cite{page}). The idea of Hawking radiation exploits the concept of creation of pair production next to the event horizon (out of the vacuum) with one of the particles running away to the infinite space from the boundary while the other with negative energy getting sucked into the black hole resulting in the decrease in its mass until the whole black hole disappears in a cloud of radiation. A natural question has been asked as to whether viable information could be gathered at temperatures near the scale of Planckian mass when the quantum gravitational effects become substantial \cite{unr}. 

In this paper our primary focus is to look at a suitable structure of a non-Hermitian two-level effective Hamiltonian \cite{volo1, volo2} to illustrate the possibility of artificial Hawking radiation. This is done by mapping to a coordinate setting and making use of the tetrad-based method. The latter strategy is frequently used for seeking solutions pertaining to a curved space. 
We will demonstrate that black hole similarities emerge with the emission of Hawking-like radiation when the event horizon causes separation of two distinct topological regions \cite{mor, sims}. 

A non-interacting field theory is often sought to address Hawking radiation. In fact, it was observed that in two dimensions Schwarzschild geometry, interaction effects are minor and that a free particle theory is adequate for the treatment of Hawking radiation, see for some detailed discussion \cite{leahy, frasca}. In the following, we adopt the procedure of Parikh and Wilczek to
estimate the tunneling probability by employing the standard classical approach of WKB approximation\cite{par}. In their formalism, effect of back-reaction was included to ensure energy conservation while a particle was emitted through the process of tunneling in going past the horizon. Noting that the need for a nonsingular coordinate system is essential at the horizon, we adopt below the well known  
Painleve-Gullstrand coordinates \cite{pain,gul} which are simply coordinate transformations of the usual Schwarzschild solution. The corresponding metric tensor, which involves an off-diagonal element, is regular at the Schwarzschild radius but has 
a singularity only at the origin, see for example \cite{kanai}. It needs to be mentioned that prior to the work of \cite{par}, the idea of Hawking radiation as tunneling was first investigated by Srinivasan and Padmanabhan \cite{sri}, including a treatment in different coordinate settings \cite{sha}.

A remark is in order: The need to use Painlev\'{e}-Gullstrand coordinates for the Schwarzschild metric arises due to the singularity in the usual Schwarzschild coordinates. However, use of Kruskal-Szekeres coordinates was shown to give an incorrect formulation \cite{akhmed1, akhmed2} implying that one should be able to use other coordinates so long as one did not have a real singularity instead of coordinate singularity.

\section{Biorthogonality and exceptional point}

The most general $2 \times 2$ Hermitian Hamiltonian (with Fermi velocity set to  unity) in terms of the linear node $\vec{k} = (k_x, k_y, k_z)$, has the form \cite{oku, arm} $\hat{h}_0 = \chi (k) + \sum_{i=x, y, z} d_i (k) \sigma_i$, where $\chi$ and the $d_i$'s are suitable real coefficients, $k = |\vec{k}|$, and $\sigma_i$'s are Pauli matrices. The Hamiltonian which characterizes conduction and valence bands supports a pair of eigenvalues whose separation is removed when the coefficients vanish simultaneously. This is at once obvious from the energy dispersion providing the energy splitting

\begin{equation} \label{eq2}
\epsilon_{\pm} = \chi \pm \sqrt{d_x^2 + d_y^2 + d_z^2}
\end{equation}
We see that degeneracy requires all three $d_i$'s to become zero together i.e. $d_x = d_y = d_z = 0$, which is not symmetry protected. In Weyl semimetals, in which the conduction and valence bands energy coincide over a certain region
of the Brillouin zone, a linear crossing of two bands takes place at the forming of nondegenerate Dirac cones corresponding to the conduction and valance bands. 
 
Extension of topological phases from the Hermitian to the non-Hermitian sector has been pursued in a variety of papers \cite{hongwu} and the presence of
gains and losses investigated \cite{yang1, yang2}. A point was made a few years ago about the question of whether real black holes can emit Hawking radiation and whether meaningful information can be gathered about Planckian physics \cite{unr}.
Very recently, De Beule et al \cite{chr} made an explicit analysis of the existence of artificial event horizon in Weyl semimetal heterostructures. In their work, the electronic analogs of stimulated Hawking
emission was studied and physical observables were identified.  Sabsovich et al \cite{sab} examined black and white hole analogs in Weyl semimetals subjected to inhomogeneous nodal tilts and explored experimentally viable consequences and showed the general relativity analogy of such Hamiltonians. Analogy was also drawn in some papers between the low-energy
Hamiltonian of tilted nodes and black hole metric \cite{volo2, zhang, solu}. In a somewhat similar context, the possibility of the emission of Hawking
radiation was looked into \cite{volo4}.
Further, imitation of black hole Hawking radiation was found in a purely classical-mechanical system by employing a coupled double chain model admitting frequency dispersion \cite{jana}. The tied up issue of the tunneling probability across the
event horizon points was explored in the framework of a two-level non-Hermitian topologically insulated Weyl-type Hamiltonian which contained titing in one of the directions \cite{bsen}. 

A two-dimensional structure of a non-Hermitian dissipative Hamiltonian also featured in an elaborate investigation of analogue Schwarzschild black holes emitting Hawking radiation \cite{sta}. A typical situation is described by the inclusion of a non-Hermitian term $i \vec{\tau}(k) \cdot \vec{\sigma}$, with $\vec{\tau} = (\tau_x, \tau_y, \tau_z)$, to $\hat{h}_0$ when the overall Hamiltonian assumes the form $\hat{h} = \hat{h}_0 + i \vec{\tau}(k)\cdot\vec{\sigma}$. In such a situation the eigenvalues turn out to be

\begin{equation}\label{eq3}
E_{\pm}(k) = \chi \pm \sqrt{\Delta}
\end{equation}
where $\Delta = d_x^2+d_y^2 + d_z^2 -(\tau_x^2+\tau_y^2+\tau_z^2) + 2i (\tau_x d_x +\tau_y d_y +\tau_z d_z)$.  A consequence is that these become equal when $\Delta = 0$ pointing to the presence of an exceptional point \cite{kato, heiss1, heiss2, zno, ply} where we have 

\begin{equation}\label{eq3.1}
d_x^2 + d_y^2+ d_z^2 = \tau_x^2+\tau_y^2+\tau_z^2 \quad \mbox{and} \quad
\tau_x d_x +\tau_y d_y +\tau_z d_z = 0
\end{equation}
Defining the right and left eigenstates to be

 \begin{gather}\label{eq4.1}
   \hat{h} (k)|\psi_{\pm}^R \rangle = E_{\pm} |\psi_{\pm}^R \rangle   \\
 \hat{h}^\dagger (k)|\psi_{\pm}^L \rangle = E^*_{\pm} |\psi_{\pm}^L \rangle 
\end{gather}
where $|\psi_{\pm}^R \rangle$ and $|\psi_{\pm}^R \rangle$ are explicitly

\begin{gather}\label{eq4.2}
   |\psi_{\pm}^R \rangle= \frac{1}{\sqrt{2(E_\pm-\chi)(E_\pm-\chi + d_z+i\tau_z)}}\left[E_\pm-\chi+d_z+ i\tau_z, \quad (d_x + i d_y)-(\tau_y-i\tau_x)\right]^T \\
   | \psi_{\pm}^L \rangle= \frac{1}{\sqrt{2(E^*_\pm-\chi)(E^*_\pm-\chi+d_z-i\tau_z)}}\left[E^*_\pm-\chi +d_z- i\tau_z, \quad (d_x + i d_y)+(\tau_y-i\tau_x)\right]^T
\end{gather}
the biorthogonality relations are a ready outcome

\begin{gather}\label{eq4.3}
   \langle \psi_{\alpha}^L | \psi_{\beta}^R \rangle =\delta_{\alpha\beta} \quad \alpha,\beta=\pm
\end{gather}
The self-orthogonality of the eigenstates can be worked out easily \cite{cerjan}.


\section{A two-level PT-symmetric model}

\subsection{The Hamiltonian}

Consider the following arrangement of the $d$-coefficients to enquire into the spectral phase transition as the system transits to exhibiting complex eigenvalues from the real ones

\begin{equation}\label{eq6}
d_x = \rho(k) \cos\phi (k), \quad
d_y = \rho(k) \sin\phi (k), \quad d_z = 0, \quad \tau_x = \tau_y = 0, \quad \tau_z = \lambda \eta(k)
\end{equation}
where k is a real variable, $\rho(k), \eta(k)$ and $\phi (k)$ are a set of real, nonzero periodic functions, and $\lambda \in {\Re}^+$ is a coupling parameter. Inclusion of the latter implies introduction of gain and loss
in the system thereby signaling the possibility of the appearance
of exceptional points where abrupt phase transitions could occur. The class of representations (\ref{eq6}) have also been studied in \cite{liang, longhi}.

Expressed in the matrix form the Hamiltonian corresponding to (\ref{eq6}) reads

\begin{equation}\label{eq6.1}
 \hat{h} \rightarrow   \hat{\mathcal{H}} = \begin{pmatrix}
        T (k) & S (k) \\[6pt]
        S^* (k) & T^* (k)
    \end{pmatrix}, \quad \hat{\mathcal{H}} \neq \hat{\mathcal{H}}^\dagger
\end{equation}
where $T = \chi(k) + i \lambda \eta(k)$ and $S= \rho(k) e^{i \phi(k)}$. The Hermitian counterpart of $\hat{\mathcal{H}} (k)$ corresponds to $k=0$. Enforcing $\mathcal{PT}$-symmetry \cite{bender1} requires the diagonal elements of the matrix generated by (\ref{eq6.1}) to be complex conjugate of each other and similarly for the off-diagonal elements too. $\hat{\mathcal{H}} (k)$ is easily seen to commute with the joint operator $\mathcal{PT}$ i.e. $[\hat{\mathcal{H}},\mathcal{PT}]=0$ with the $\mathcal{P}$ operator represented by $\sigma_x$,
and $\mathcal{T}$ standing for the usual complex conjugation operation. 

Apart from analytical evaluations \cite{bender2}, numerical algorithms for the diagonalization of $\mathcal{PT}$-symmetric Hamiltonians have also been carried out in the literature \cite{noble2}. The possibility of $\mathcal{PT}$-symmetry residing in quantum mechanical systems is a prominent forefront in research. Their position was soon found out to be intermediate between open and closed systems. While the role of non-Hermiticity in understanding stable phases has been pursued in the literature for last several years \cite{wei, yuce}, the character of $\mathcal{PT}$-symmetry for stable nodal points concerning gapped and gapless semimetals \cite{goe, arm}, where the invariants are constituted by 
Bloch bands, is a somewhat recent realisation. Indeed due to such a symmetry prevailing, one finds stable nodal points to exist in lesser
dimensions \cite{ber}.

The eigenvalues of $\hat{\mathcal{H}}$ are easily seen to satisfy the relation 

\begin{equation}\label{eq7}
\mathcal{E}_{\pm}(k) = \chi(k) \pm \eta(k)\sqrt{\overline{\lambda}^2 (k) -\lambda^2}, \quad \overline{\lambda} (k) \equiv \frac{\rho(k)}{\eta(k)}
\end{equation}   
Introducing $\tan \theta = \frac{\rho(k)}{i\lambda \eta(k)}$ where $\theta = \theta (k)$, one can classify the accompanying right eigenvectors to be 
    \begin{equation}\label{eq8}
 \left [ \cos (\frac{\theta}{2}),\, \sin (\frac{\theta}{2}) e^{-i\phi} \right ]^T,  
   \;
       \left [ \sin (\frac{\theta}{2}),\,  -\cos (\frac{\theta}{2}) e^{-i\phi} \right ]^T
 \end{equation}
while their left partners are
\begin{equation}\label{eq9}
 \left [ \cos^* (\frac{\theta}{2}), \,\sin^* (\frac{\theta}{2}) e^{-i\phi} \right ]^T,
 \;
     \left [ \sin^* (\frac{\theta}{2}),\,  -\cos^* (\frac{\theta}{2}) e^{-i\phi} \right ]^T
 \end{equation}
It can be readily checked that these obey the biorthogonal conditions (\ref{eq4.3}).

The expression (\ref{eq7}) clearly shows that the eigenvalues stay real when the inequality $\lambda < \overline{\lambda}$ holds. This corresponds to the situation when $\mathcal{PT}$ is unbroken. However, when opposite is the case, i.e. $\lambda > \overline{\lambda}$, a broken $\mathcal{PT}$ phase is encountered. At the critical value $\lambda = \overline{\lambda}$, the exceptional points appear when both the eigenvalues $\mathcal{E}_+$ and $\mathcal{E}_-$ coincide to become $\chi(k)$ and the associated eigenvectors coalesce to form a single entity. In other words, at the exceptional points we have $\rho = \pm \lambda \eta$. 

It is worthwhile to mention that a simpler form of $\hat{\mathcal{H}} $, proposed by Bender et al \cite{bend} a few years ago, discusses the type $\hat{\mathcal{H}} =q\cos\phi \P +iq\sin\phi \sigma_z +s\sigma_x$ which supports the eigenvalues $\lambda_{\pm} = q\cos\phi \pm\sqrt{s^2-q^2\sin^2\phi}$. These remain entirely real when the inequality $s^2>q^2\sin^2\phi$ is obeyed.

\subsection{The tetrad representation}
 
Let us represent the Hamiltonian in the following tetrad basis

\begin{equation}\label{eq10}
    \hat{\mathcal{H}}  = e{^\mu}{_a} h_\mu\sigma^a+e{^\mu}{_0} h_\mu\P
\end{equation}
and $h_\mu$'s are suitable entities to be matched. The vielbiens $e{^\mu}{_a}$ and $e{^\mu}{_0}$ satisfy the orthonormality condition $e{^a}_\mu e{^\mu}_b=\delta{^a}_{b}$, $\mu = (0,x,y,z)$ and $a, b = (x,y,z)$ subject to the metric being expressed as the bilinear combination of the tetrads

\begin{equation}\label{eq11}
    g^{\mu\nu}=e{^\mu}{_\alpha} e{^\nu}{_\beta}\, \eta^{\alpha \beta}
\end{equation}
where $ \eta^{\alpha \beta} = diag(-1,1,1,1)$ is the Minkowski metric
of flat spacetime.  

To proceed with (\ref{eq11}), let us first write the vielbiens $e_a$${^\mu}$ in terms of four functions $f_1, f_2, g_1, g_2$, and then seek to relate them with the known functions at hand namely, $\chi(k), \rho(k), \eta(k), \phi(k)$. A convenient set of vielbiens is given by \cite{kim}

\begin{equation}\label{eq12}
e_0 {^0} = f_1, \quad e_1 {^0} = f_2, \quad e_0 {^1} = g_2, \quad e_1 {^1} = g_1, \quad
e_2 {^2} = \frac{1}{r}, \quad  e_3 {^3} = \frac{1}{r\sin\theta}
\end{equation}
with the respective inverses 
\begin{gather}\label{eq13}
e^0{_0} = \frac{g_1}{f_1g_1 - f_2g_2}, \quad e^0{_1} = -\frac{f_2}{f_1g_1 - f_2g_2}, \quad e^1{_0} =  -\frac{g_2}{f_1g_1 - f_2g_2}, \quad \\ \nonumber
e^1{_1} = \frac{f_1}{f_1g_1 - f_2g_2},\quad e^2{_2} = r, \quad  e^3{_3} = r\sin\theta
\end{gather}
The line element then simplifies to the form

\begin{equation}\label{eq14}
ds^2 = \frac{g_1^2 -g_2^2}{f_1g_1^2} dt^2 + \frac{2g_2}{f_1g_1^2} dt dr -\frac{1}{g_1^2} dr^2 - r^2 d\Omega^2
\end{equation}
The Schwarzschild gauge arises for the following choice  of $f_1, f_2, g_1, g_2$ namely

\begin{equation}\label{eq15}
f_1 = 1, \quad  f_2 = 0, \quad g_1 = 1, \quad g_2 = - \sqrt{\frac{2\mathcal{M}}{r}}
\end{equation}
and yields the following form of the black hole metric in Painlev\'{e}-Gullstrand
coordinates

\begin{equation}\label{eq16}
    ds^2 = \left (1-\frac{2\mathcal{M}}{r}\right ) dt^2 - 2 \sqrt{\frac{2\mathcal{M}}{r}} dr dt - dr^2 -r^2 d\Omega^2
\end{equation}
where $\mathcal{M}$ is the mass of the black hole and $t$ represents the Painlev\'{e} time. The metric (\ref{eq16}) is stationary (i.e. invariant under translation of $t$) but not static (i.e. not invariant under time-reversal) and is consistent with the transformation proposed originally in \cite{pain, gul}. 

With the help of (\ref{eq10}) and (\ref{eq12}), the general form of the Hamiltonian emerges as
\begin{equation}\label{eq17}
\begin{split}
\hat{\mathcal{H}}  = \frac{g_1h_0 - g_2 h_1}{f_1g_1 - f_2g_2} \P + \frac{-f_2 h_0 + f_1 h_1}{f_1g_1 - f_2g_2} \sigma_x 
 + rh_2 \sigma_y + r\sin\theta h_3 \sigma_z
\end{split}
\end{equation}
Comparing with (\ref{eq14}), we can easily derive the coordinate-space correspondence

\begin{equation}\label{eq18}
\begin{split}
\chi(k) \rightarrow \frac{g_1h_0 - g_2 h_1}{f_1g_1 - f_2g_2}, \quad \rho(k)\cos\phi \rightarrow \frac{-f_2 h_0 + f_1 h_1}{f_1g_1 - f_2g_2}, \quad \rho(k)\sin\phi \rightarrow r, \quad \lambda \eta(k) \rightarrow r\sin\theta 
\end{split}    
\end{equation}
where we have specified $h_0 = h_1 = h_2 = 1$ and $h_3 = i$. Using the values in (\ref{eq15}), we have the mapping correspondence to the $(r, \theta)$ variables

\begin{equation}\label{eq19}
\begin{split}
\chi \rightarrow 1 + \sqrt{\frac{2\mathcal{M}}{r}}, \quad \rho\cos\phi \rightarrow 1, \quad \rho\sin\phi \rightarrow r, \quad \lambda\eta \rightarrow r\sin\theta 
\end{split}
\end{equation}
which is consistent with $\rho \rightarrow \sqrt{r^2 + 1}$ and $\phi \rightarrow \tan^{-1} (r)$. As a result, the Hamiltonian assumes the form

\begin{equation}\label{eq20}
    \hat{\mathcal{H}}  = \left (1+\sqrt{\frac{2\mathcal{M}}{r}} \right ) \P + \sigma_x + r \sigma_y + ir\sin\theta \sigma_z
\end{equation} 
In the following we estimate the probability transmission amplitude of the analogue Hawking radiation by making use of the correspondence set up in (\ref{eq19}).

\section{Analogue Hawking radiation and tunneling estimate}

First of all, the energy eigenvalues acquire read

\begin{equation}\label{eq21}
    \mathcal{E}_{\pm} = \left (1 + \sqrt{\frac{2\mathcal{M}}{r}} \right ) \pm \sqrt{1 + r^2 \cos^2 \theta}
\end{equation}
on using (\ref{eq7}). The exceptional points correspond to $r = \pm i \sec\theta$ located on the imaginary axis. In what follows we will adhere to the positive sign. We then have

\begin{equation}\label{eq22}
    d\mathcal{E} = \frac{d\mathcal{M}}{\sqrt{2\mathcal{M}r}}
\end{equation}
Before we calculate the tunneling probability, let us note that when the particle escapes from the black hole with an energy $\omega$, the mass of black hole decreases from $\mathcal{M}$ to $\mathcal{M}-\omega$. Indeed, as the pair production takes place around the event horizon, the positive energy particle when breaks free \cite{par} has to transit the separating region defined between $r_{in}$, which is the radius of the black hole before the emission of the particle, and $r_{out}$, which is the radius of the black hole after the emission of the particle, acting as a possible barrier wall. This is possible if the particle can tunnel through such a barrier. Actually, a classically inaccessible zone is replicated with the particle possessing energy below such a resistance. 

In dealing with the tunneling problem we observe that since the action $\zeta$ in the transmission region is imaginary, the probability of tunneling to take place can be straightforwardly calculated by making use of the semiclassical WKB approximation \cite{par}. Here an s-wave particle is considered to go outwards from $r_{in}$ to $r_{out}$ meaning that $\zeta$ can be cast in the form

\begin{equation}\label{eq23}
\mbox{Im}\zeta=\mbox{Im}\int_{r_{in}}^{r_{out}}p_rdr = \mbox{Im}\int_{r_{in}}^{r_{out}}\int_{0}^{p_r}dp_rdr
\end{equation}
where the equation of motion for the canonical momentum  
is imposed as prescribed  by the classical Hamilton's equation. Noting that the Hamiltonian assumes the respective values $\mathcal{M}$ and $\mathcal{M} - \omega$ for $p_r=0$  and $p_r=p_r$, $\mbox{Im}\zeta$ can be transformed to \cite{bag2}

\begin{equation}\label{eq24}
    \mbox{Im}\zeta = - \mbox{Im}\int_{r_{in}}^{r_{out}}\int_{0}^{\omega}\displaystyle\frac{d\omega}{\dot{r}}dr
\end{equation}
where, since the emitted particle has a very negligible mass, we can approximate 
$\mathcal{M} = \mathcal{M}_{\mbox{in}} \approx \mathcal{M}_{\mbox{out}}$ and
        $d \mathcal{M}  \approx -d\omega $.
    
    For the metric at hand, the presence of the horizon can be determined from the radial null geodesic condition $ds^2 = 0$ corresponding to (\ref{eq16}). The resulting differential equation becomes

\begin{equation}\label{eq25}
    \dot{r}^2 + 2 \sqrt{\frac{2\mathcal{M}}{r}} \dot{r} - \left(1 - \frac{2\mathcal{M}}{r}\right) = 0
\end{equation}
which admits of the following acceptable solution

\begin{equation}\label{eq26}
    \dot{r} = 1 - \sqrt{\frac{2\mathcal{M}}{r}}
\end{equation}
Substituting in (\ref{eq24}) results in

\begin{equation}\label{eq27}
    \mbox{Im}\zeta= - \mbox{Im}\int_{2\mathcal{M}}^{2(\mathcal{M}-\omega)}\int_{0}^{\omega}\frac{d\omega dr}{1-\sqrt{\frac{2(\mathcal{M}-\omega)}{r}}}
\end{equation}
which can be reduced to a tractable form on using the residue theorem with the help of the substitution $\alpha^2 = \frac{2(\mathcal{M}-\omega)}{r}$. This yields

\begin{equation}\label{eq28}
    \mbox{Im}\zeta= - \mbox{Im}\int_{2\mathcal{M}}^{2(\mathcal{M}-\omega)}dr\int_{0}^{\omega}\frac{d\alpha}{1-\alpha} = 4\pi\omega
\end{equation}
Extracting the imaginary part from the right side, we obtain for the tunneling probability the result
\begin{equation}\label{eq29}
\Gamma \approx e^{-2\,\mbox{Im}\zeta} = e^{-8\pi\omega}
\end{equation}
A similar estimate was made in \cite{par} on accounting for the global conservation laws. It is to be noticed that $\mbox{Im}\zeta$ as given by (\ref{eq29}) is independent of the parameters of the model Hamiltonian. It is also important to point out that we do not claim invariance under canonical transformation of the tunneling rate as given by (\ref{eq16}) as also follows from the works of \cite{akhmed1, akhmed2}. Furthermore, for the inclusion of the temporal contribution to the tunneling rate we refer to the works in \cite{akhmedo, akhmed3}. 

In conclusion, concerning the observable that could be associated with the Hawking radiation, a natural question that arises as to what is the observable that would result analogous to the Hawking radiation and in particular, the decay rate of complex eigenstates. Indeed, because of the near impossibility to observe Hawking radiation in a real black hole \cite{cast}, seeking black-hole analogues has become an interesting alternative because these could reveal properties akin to gravitational black holes, emission of Hawking-like radiation being a specific one. In this regard, construction of experimental set-ups has been undertaken to identify quantum fluctuations that might emerge \cite{stein}. Also, very recently, observation of stimulated Hawking radiation was reported which occurs in a regime of extreme nonlinear fiber optics \cite{drori}. Another point which requires a more detailed study is the relationship between the decay rates addressed in \cite{dente, bus, rotter} and the one estimated in (\ref{eq29}). As a final remark, let us state that we can formally extend our model to other black hole metrics like the charged black holes in 2D dilaton gravity which originates from the low energy effective theory of type 0A string theory \cite{kim} as well as for the rotating and charged black hole background \cite{umetsu}.


\section{Summary}

Non-Hermitian Hamiltonians are found to play a central role in diverse physical problems. In this paper we studied a generalized form of a two-level $\mathcal{PT}$-symmetric system that depends on a real parameter k and exhibits a period of $2\pi$. Adopting the approach of tetrad formalism a correspondence was established between such a Hamiltonian and the one constructed in terms of vielbeins. This enabled us to connect to the metric of curved spacetime. By suitably writing the tetrad components in
terms of four unknown functions and making a specific choice of them such that the Schwrarzschild metric could be described in Painlev\'{e}-Gullstrand
coordinates, we could put in a one to one correspondence with our chosen form of the Hamiltonian. We computed the probability transmission amplitude  of Hawking radiation by looking at it as a tunneling process and making use of the semiclassical WKB approximation. Our result turned out to be independent of the non-Hermitian parameter $\lambda$ so that the nature of phase transitions that the system supports does not influence the tunneling probability estimate.


\vspace{6pt} 




\section{Data Availability}
All data supporting the findings of this study are included in the article.

\section{Acknowledgments}
We thank all the anonymous referees for making a number of constructive remarks. We also thank Andreas Fring for pointing a couple of errors in an earlier version of the manuscript. Two of us (RG, SS) thank Shiv Nadar IoE for the grant of research fellowships.

\section{Conflicts of Interest}
The authors declare no conflict of interest.











 \end{document}